# DATA ORIENTED ALGORITHM FOR REAL-TIME ESTIMATION OF FLOW RATES AND FLOW DIRECTIONS IN A WATER DISTRIBUTION NETWORK

Christophe Dumora[1,2,3] & David Auber[1] & Jérémie Bigot[2] & Vincent Couallier[2] & Cyril Leclerc[3]

[1]*Laboratoire Bordelais de Recherche en Informatique, Université de Bordeaux, Talence, France*
[2]*Institut de Mathématiques de Bordeaux, Université de Bordeaux, Talence, France*
[3]*LyRE, Centre de R&D SUEZ, Talence, France*
email: christophe.dumora@u-bordeaux.fr

**ABSTRACT**

The aim of this paper is to present how data collected from a water distribution network (WDN) can be used to reconstruct flow rate and flow direction all over the network to enhance knowledge and detection of unforeseen events. The methodological approach consists in modeling the WDN and all available sensor data related to the management of such a network in the form of a flow network graph $G = (V, E, s, t, c)$, with $V$ a set of nodes, $E$ a set of edges whose elements are ordered pairs of distinct nodes, $s$ a source node, $t$ a sink node and $c$ a capacity function on edges. Our objective is to reconstruct a real valued function $f(u,v): V \times V \Rightarrow \mathbb{R}$ on all the edges $E \in V \times V$ from partial observations on a small number of nodes $V = \{1, ..., N_V\}$. This reconstruction method consists in a data-driven Ford-Fulkerson maximum-flow problem in a multi-source, multi-sink context using a constrained bidirectional breadth-first search based on Edmonds-Karp method. The innovative approach is its application in the context of smart cities to operate from sensor data, structural data from geographical information system (GIS) and consumption estimates.

## 1. INTRODUCTION

As reported by the World Health Organization, the urbanization rate (UR) in 2014 accounted 54% of the global population, up from 34% in 1960, and continues to grow. This UR is expected to be slightly above 60% in 2030 (United Nations, 2007). The planet today counts 3.3 billion city dweller, which is four and a half times more than in 1950. By 2030, the urban population is expected to reach 5 billion; there would be as many city dwellers in the world as inhabitants on Earth in 1987 (Veron, 2007). In this context, the concept of smart city has recently been introduced to emphasize the importance of information and communication technologies (ICT). As cities become more attractive and welcoming, it is necessary to be able to understand and master the complex interactions between technological and service delivery systems (Caragliu et al., 2011). The analysis of data collected from sensors and their connectivity offer the possibility to tackle these novel issues. Distributing water that complies with sanitary requirements while providing an irreproachable taste is one of the fundamental services offered by a city to his citizens.

A smart water distribution network (WDN) must ensure the supply of safe drinking water from its extraction to different raw water sources, to the taps of consumers, while controlling the impact on both hydraulic and energetic resources. This task is based on the good management of hydraulic parameters (such as flow rate or operating pressure), on the control of water quality through treatment by purification (such as chlorination), but also by controlling water demand. For these reasons, WDN are instrumented with sensors. Positioning sensors on the production plants, on specific point of the WDN, or closer to the consumer make it possible to monitor and to control key parameters related to water quality (e.g. chlorine, pH, temperature) or the water demand (e.g. consumption). However, just as cities are expanding, WDN are also growing. WDN are increasingly meshed to ensure better security of supply. This looped pipe configuration greatly complicates the hydraulic and increase the risk of stagnation which represent a critic situation that must be avoid prevent water pollution (Obrodovic et al., 1998; Manual of Water Supply Practices, 2005).

The information provided by the sensors therefore leaves shadow areas in which it is difficult to control all the phenomenon taking place on a small spatial and temporal mesh.
One of the main challenges for the control and management of WDN is to be able to estimate the key parameters at any point in the network to remove these shadows areas.

The contribution lies in the methodological approach that consists in modeling a WDN and all available sensor data related to the management of such a network in the form of a network graph $G = (V, E)$. The graph structure use a set nodes $E$, edges $V$ and properties to represent and store data. This design allows simple and fast retrieval of complex hierarchical structures that are habitually difficult to model when data are heterogeneous. Our objective is to reconstruct a vector of numerical characteristics on all edges $E \in V \times V$ from partial observations on a small number of nodes $V = \{1, ..., N_V\}$. This reconstruction method consists in a data-driven maximum-flow algorithm in a multi-source, multi-sinks context. We experimented the method to infer sensors data throughout the Bordeaux Metropolis's WDN in France. With more than 3000 km of pipes this network delivers an average of 170 000 m$^3$ of water per day for more than 700 000 consumers. The paper is organized as follows. First, we describe challenges related to the management of WDN in the context of smart cities. Second, we present the data generated by the operation of a WDN required to build a flow network. We then describe a constrained Ford-Fulkerson method to estimate flow direction and flow rate inside a WDN. We conclude by a discussion about the future work that will be done to enhance the methodology and results.

## 2. SMART WATER CHALLENGES

Most of the time the small number of sensors do not allow to obtain sufficient information at every point of the network. Hydraulic models are therefore used to provide complete information on the hydraulic operation of the network (travel times, flow rates, etc.). Nevertheless, these models, although accurate, become more and more complex to solve. They require the use of average consumption profiles and typical configurations (peak days, winter, summer, etc.). This makes them become less adapted to determine, in real time, the state of the network. Estimating the hydraulic condition of the WDN on every point and time meets many challenges. The United Nations Organization estimated that, by 2025, two-thirds of the world's population could live in water stress conditions. The increasing world's population, climate changes, growth in agricultural production and access to drinking water play a significant role in the increased demand. Furthermore, even if in the last few years, the consumer tendency declines, the water demand continues to increase. Future water shortages will require increased efficiency in the treatment and transmission of water. For example, real time flow direction knowledge, would allow us to visualize the water origins at any specific point and time in a WDN. In this way, we could precisely know the path traveled by the water and track all events or interventions that could have an impact on its quality.

This information obtained in real time, could play a key role in improving the management of the WDN, including:
- water shortage risk identification

- water extraction optimization
- pollution impact estimation
- real time water quality monitoring
- enhanced leaks identification
- risk areas prediction (such as red water)
- anomaly detection
- etc.

Thus, the WDN environment must be considered as a whole to get a smarter water management but also more resilience in continuity and quality supply. The interactions between ICTS (Information and Communication Technologies) and WDN structural data need to be further considered jointly. The way we chose to consider this interconnectivity and interdependence associated to the WDN is as a graph network.

## 3. GRAPH STRUCTURE

Many data initially disconnected from each other can be regrouped and linked together on the WDN. As shown on Figure 1 these data are initially regrouped by theme. This can range from water quality data (e.g. punctual or automatic measure of quality parameters, quality consumer complaint), to field intervention data (e.g. renewal, leak repair, valve actuation), through hydraulics information (e.g. volume delivered to the network, hydraulic simulations, consumed volumes). This set of heterogeneous data, because of their formats, uses and locations, can nevertheless be interconnected to each other at all levels thanks to the graph network. It is important to consider all these data, their interconnectedness and their interdependences to be able to understand the phenomenon that can disrupt the functioning of the network. We propose a more general data structure which is the network graph model. This structure well studied in the field of graph theory, is adapted to intensive calculation and data visualization.

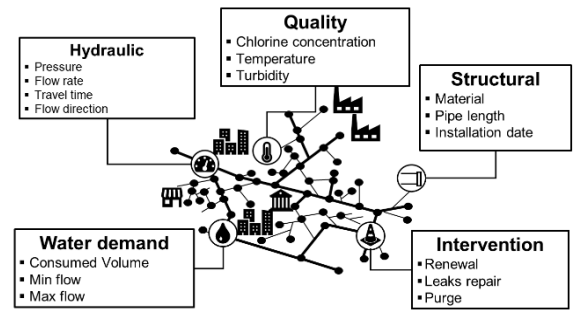

Figure 1. Schematic data representation related to the management of water distribution network.

In this part we will focus on two types of information attached to the nodes and edges of the graph, to build a flow network on which we want to apply a flow reconstruction method. 3.1) the structural data which represent the skeleton of the WDN and provide the nodes, edges objects, 3.2) the sensors data which concern the monitoring of all measures related to the flow. We present in section 3.3 how we build a flow network according to the data collected to perform the maximum flow problem.

### 3.1 Structural Data

They represent each of the structural objects of the WDN (e.g. water production plant, pipes, connectors, valves, sensors). All these objects are geo-referenced and described by characteristics such as pipe length, pipes diameter, state of valves etc. These data represent the skeleton of the network. Beyond representing the set of structural data, they have the central role of providing a structure on which all data related to the WDN can be linked.

We now consider the WDN as a graph network $G = (V, E)$, where $V = \{1, \ldots, N_V\}$ is the set of nodes and $E \in V \times V$ the set of edges. Each node represents a structural object of the WDN. Each edge depicts physical link between the nodes in our case edges represent pipes. If the nodes $u$ and $v$ are linked by a pipe we create and edge $E_{u,v}$, of length $l_{u,v}$ and internal diameter $d_{u,v}$.

The structural properties of pipes allow us to determine a theoretical maximum possible flow for each pipe depending on their diameters and a maximum flow speed fixed. We define $Cmax_{(u,v)}$ as the capacity of the pipe $E_{u,v}$ representing the maximum possible flow in $m^3/h$ for the edge $E_{u,v}$.

$$Cmax_{(u,v)} = s_{(u,v)} \times V_{max} \qquad (1)$$

With $s_{u,v} = \pi \times \left(\frac{d_{u,v}}{2}\right)^2$ the surface and $V_{max}$ the maximum flow speed of the pipe $E_{u,v}$.

### 3.2 Sensors

In order to monitor the key parameters related to the management of the WDN, sensors are installed at different strategic points. These sensors provide local information about water that cross through the network on small time step. In the context of building a flow network the sensors in which we will focused are counting devices providing flow direction and flow rate. These flows information are not fixed. They can vary a lot inside the same pipe. Flow direction is not necessary obvious to determine and can reverse depending on the period and the total volume consumed.

These sensors are crucial to obtain real information of the flow rates and flow directions. They measure the amount of water passing through a specific point of the network in $m^3/h$, with a time-frequency around 1 measure per 15 minutes. We consider in this paper two types of sensors nodes:

- Type a: Those providing inflow or outflow information. Divided into two categories, sources nodes $V_s \forall s \in V_s = \{1, ..., n\}$ and sinks nodes $V_t \forall t \in V_t = \{1, ..., m\}$, depending if it measures inflow or outflow.
  The sources nodes are mainly located on reservoirs or at the water production plants outputs. They measure the amount of water delivered on the network. The sinks ones can be located on interconnections with other townships networks to provide water in case of shortage or on water tanks when they are filled by water production plant, but especially at consumers level, thanks to Automatic Meter Reading (AMR), counting on an infra-day period the quantities of water consumed. Inflow and outflow information are crucial to be able to reconstruct flow inside a network. It provides quantity and position information of water entering or leaving the network.
  But not all water meters are equipped with AMR. We need to estimate the unknown consumption of the non-AMR to be able to apply the maximum flow problem. These non-AMR consumptions could be estimated in many ways, from smoothing an equal proportion of the missing outflow at all the unknown sinks, to an individual consumption estimate (Ghul & Brémond, 2000). In this paper we used a methodology presented in Claudio et al. (2014) which was the one currently used by Suez who deals with the case study contract. This methodology consists in an annual consumption level group division of the population based on a Horvitz-Thompson estimator. The non-AMR consumption is then estimated according to the average AMR consumption of the group in which they belong.

- Type b: Those providing observations within the network. These sensors measure the quantity of water passing through the node giving local information about the flow rate and flow direction. According to sensors nodes we can constrain the capacity $Cmax_{(u,v)}$, introduced earlier in section 3.1, with $x_{obs}$ the value measured by the sensor. So, for all edges connected to a sensor node in $V_{obs} = \{1, ..., n_{obs}\}$, the maximum possible flow become an observed flow $Cmax_{(u,v)} = x_{obs}(u) \forall u \in V_{obs}$. Moreover, the read direction of sensors is known, giving us the flow direction of the flow rate measured. We can then set the direction of the adjacent edges of any sensors.

### 3.3 Flow Network Construction

In this section we present how we build a flow network according to the data presented earlier to apply a data driven maximum flow problem in our case of a WDN.

First things first, as define in graph-theoretic field (Hieneman et al., 2008; Ahuja et al. 1993) a flow network $G = (V, E)$ must be a directed graph in which each edge $(u, v) \in E$ has a nonnegative capacity $c(u, v) \geq 0$. The nonnegative capacity was set through structural information for non-measured edges and by the sensors observations for the measured ones (section 3.1 and 3.2). But our network is initially semi-directed because all the edges are not observed, and the flow directions are not known.

Because flow could potentially come from any side of the pipe, we duplicate all the edges to create bidirectional edges in the network. For each edge $(u, v) \in E$ we add an extra edge to obtain two antiparallel edges $(u, v)$ and $(v, u)$ and we set the capacity of both edges to the original one $Cmax_{(u,v)} = Cmax_{(v,u)}$. Like in a real-world, flow inside pipe can only go in one direction at a time, so the flow is not allowed to be in both antiparallel edges in the same time. For the measured edges the flow direction is known so the extra edge created will not be used. We will constrain the algorithm to not take a path that could contradict an observed edge.

We need to distinguish two vertices type in flow networks: the sources $s$ and the sinks $t$. To determine a maximum flow each vertex must lies on some path from the source to the sink. Our flow network has several sources $|V_s| = 60$ and sinks $|V_t| = 34K$ rather than just one. As described in Cormen et al. (2009), we can reduce the problem of determining a maximum flow with multiple sources and multiple sinks to an ordinary problem (Figure 2).

To do so we add a *supersource* node $s$ and directed edges $E_{s,s_i}$ for each of the source nodes. The capacity $Cmax_{(s,s_i)}$ is set to be the measured inflow of the sensor source node $s_i$.

We also create a new *supersink* node $t$ and add directed edges $E_{(t_i,t)}$ for each of the $t_i$ sink nodes. The capacity $Cmax_{(t_i,t)}$ is set to be the measured outflow of the node $t_i$.
Intuitively, any flow in the multi-sources, multi-sinks network remains the same. The single source $s$ provides as much as flow as desired by the multiple sources $s_i$ and the single sink $t$ likewise consumes as much as flow as desired for the multiple sinks $t_i$.

We now have a flow network as a directed graph $G = (V, E, s, t, Cmax)$, where $V$ and $E$ are node set and edge set, respectively; s and t are the *supersource*, and the *supersink*, respectively; and $Cmax$ is a nonnegative capacity on the edges. Let $f(u, v)$ be a real-valued function $f: V \times V \Rightarrow \mathbb{R}$. The nonnegative quantity $f(u, v)$ is the flow from the vertex $u$ to vertex $v$. We want to reconstruct the flow $f(u, v) \ \forall u, v \in V$ in $G$ that satisfies these constraints:
- Maximum flow: $f(u, v) \leq Cmax_{(u,v)} \ \forall u, v \in V$
- Conservation : $\sum_{v \in V} f(u, v) = 0 \ \forall u \in V - \{s, t\}$
- Antiparallel coherence: $f(u, v) > 0 \Rightarrow f(v, u) = 0, \forall u, v \in V$

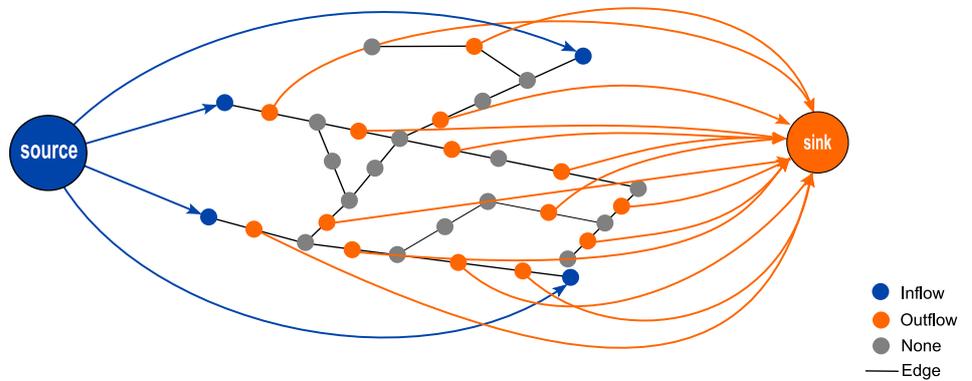

Figure 2. Converting a multiple-source, multiple-sink problem into a problem with a single source and a single sink.

## 4. FLOW RECONSTRUCTION

We choose an implementation in C++ of the Ford-Fulkerson method (Ford & Fulkerson, 1962) to compute the min-cut max-flow problem. Like in Edmonds-Karp method (Edmonds & Karp, 1972) we use a breadth-first search (BFS) algorithm to find the augmenting path. We want to constrain the algorithm to find path first by observed edges to saturate the flow in these edges. To do so we consider, $preferred = E_{(u,v)} \forall u, v \in V_{obs} \times V_{obs}$ as a vector containing all the observed edges for which the flow rates and flow direction are known. These edges are considered as *preferred* edges. Because they represent real observation in the network, we must first saturate them while solving the maximum flow problem. These *preferred* edges must be traveled in the observed flow direction. The flow inside them can only increase resulting in a constrained maximum solution but not the maximum flow solution of the Ford-Fulkerson method as described in the figure bellow.

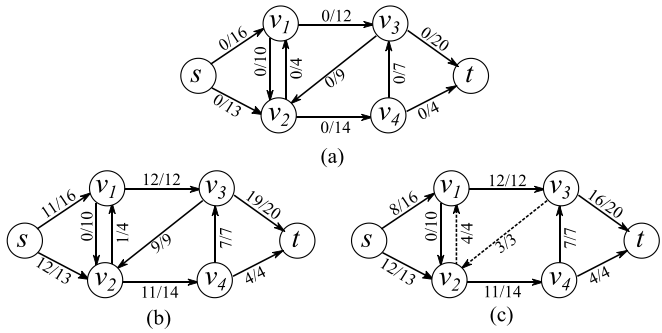

Figure 3. (a) A flow network $G = (V, E)$ with one source $s$ and one sink $t$. Each edge is labeled with its flow and capacity (flow/capacity). (b) A flow $f_1$ in $G$ with value $|f_1| = 23$ with no preferred edges. (c) A flow $f_2$ in $G$ with value $|f_2| = 20$ respecting the observed flow direction and capacity of the edges $(v_2, v_1)$ and $(v_3, v_2)$.

A bidirectional BFS is used to force path to go from the *supersource* $s$ to the *supersink* $t$ passing by all the *preferred* edges. As described in Figure 4, it starts at source node $s$ and explore the neighbors nodes before moving to the next level neighbors. While a path exists and for each nodes of the same level we repeat the same process until the sensor node is reach. The operation is repeated to find a second path starting to the sensor node to the nearest sink passing by edges still unvisited by the first path to avoid loop in the network. If the two paths are not empty, we found a valid path from $s$ to $t$ through the sensor node $s \leadsto preferred \leadsto t$.

We can add a flow $f_p = min\{c_f(u, v) \; \forall (u, v) \in p\}$ through the path representing the smallest remaining maximum capacity of all edges in the path.

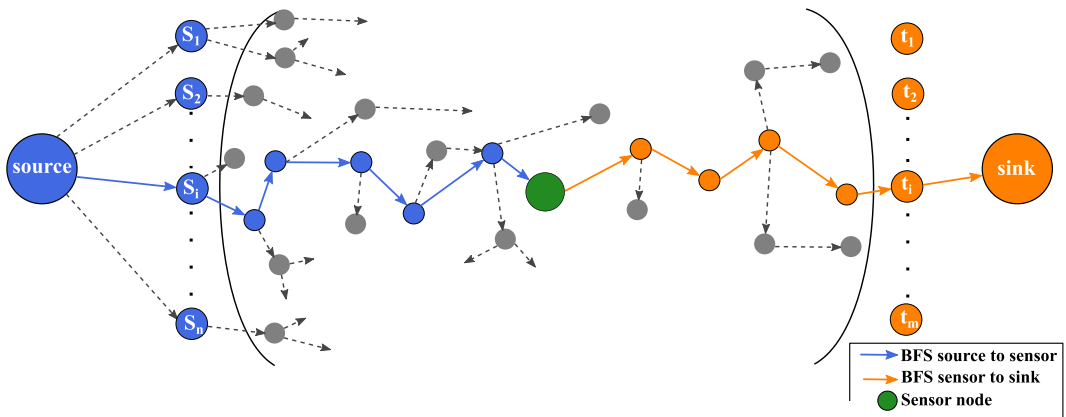

Figure 4. Bidirectional breadth-first search from source node s to sensor node v and from v to sink node t.

This process is repeated for all the *preferred* edges until there is no more path available going through them from $s$ to $t$ or if the flow of the *preferred* edges has reach the observed flow ($f(u,v) = Cmax(u,v) \forall u, v \in V_{obs}$). If the sensors capacities are reached or if there is no more augmenting path available for the sensors edges, we try to add all the remaining flow in the network as a classical maximum flow problem by looking at augmenting path from $s$ to $t$. Decreasing the maximum flow or going through reverse direction on observed edges is still forbidden during this process. For all the non-observed edges both antiparallel edges are available.

But as described in section 3.3 theses antiparallel edges represent the same physical pipe and water can't be at the same time in both. When a new flow goes through an edge in the same direction we simply add it (see line 12 Algorithm 1). When a new flow goes through an antiparallel edge while the edge was already visited in the other direction, the flow must be updated to respect the antiparallel coherence constraint and to determine in which edges the remaining flow is.

| | FORD-FULKERSON CONSTRAINED(G, s, t, *preferred*) |
|---|---|
| 1 | **while** there exists a path |
| 2 |     **if** all the observed edges are not visited or saturated |
| 3 |         $p_1 \leftarrow$ path from $s$ to *preferred* |
| 4 |         $p_2 \leftarrow$ path from *preferred* to $t$ |
| 5 |         $p \leftarrow p_1 \cup p_2$ path from s to t by *preferred* |
| 6 |     **else** $p \leftarrow$ path from $s$ to $t$ |
| 7 |     $c_f(p) = \min \{c_f(u,v), \forall (u,v) \in p\}$ |
| 8 |     **for** each edge $(u,v) \in p$ |
| 9 |         $e_0 \leftarrow$ first edge of $(u,v)$ |
| 10 |         $e_1 \leftarrow$ second edge of $(u,v)$ |
| 11 |         **if** edge in path same direction than in graph |
| 12 |             $flow(u,v) \leftarrow flow(u,v) + c_f(p)$ |
| 13 |         **else** |
| 14 |             $flow(v,u) \leftarrow flow(v,u) + \max(0, c_f + flow(u,v))$ |
| 15 |             $flow(u,v) \leftarrow \max(0, flow(u,v) - c_f(p))$ |

Algorithm 1 Pseudo code of the constrained Ford-Fulkerson method.

The constrained Ford-Fulkerson algorithm presented in the pseudo code Algorithm 1 returns for all edges a flow rate. For all the non-observed edges the flow direction is known according to the nonnegative edges. For all the observed edges we have the flow rate regarding to the expected flow. This information will help us determine if the maximum flow solution is close to observed state of the network. The Figure 5 present an output of the Algorithm 1.

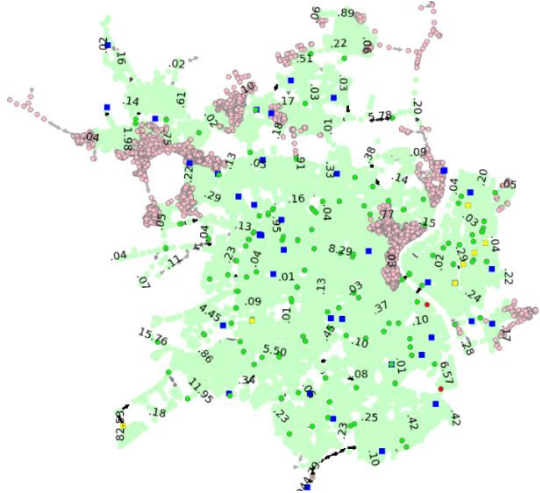

Figure 5. Output of the Algorithm 1 on the Bordeaux Metropolis (France) WDN composed by $N_V = 60K$ nodes and $N_E = 66K$ edges. In light green sink nodes that have been saturated, in light red non-saturated sinks. Squares represent sources nodes. Saturated ones are in blue and those with remaining flow are in yellow. The green dot represents sensor nodes saturated during the bidirectional BFS, and the red ones are those where the flow didn't reach the measured flow.

# 5. CONCLUSION AND DISCUSSION

We presented a methodology to create a flow network to estimate flow directions and flow rates inside a real WDN from sensors and GIS data. These results allow us to determine the age and origin of water in any point and time of the network which is an important result to answer to most of the future WDN operation challenges. We introduced a constrained Ford-Fulkerson algorithm to take advantage of all the sensors information and force the solution to respect the observed value. The result obtained on the case study of the Bordeaux Metropolis WDN are satisfying. They helped us to detect inaccurate information in both structural and sensors data. We could for example check the validity of the flow direction given by a sensor, if no flow was able to go through it during the algorithm and if by reverting it we were able to saturate the edge. The information was then confirmed by a real observation and updated into the database.

The fact that the outflow information isn't completely known on the case study do not allow to saturate all the sensors nodes. Indeed, more than 50% of the outflow is estimated. Since this estimate is based on annual consumption profile, there is area of the network where consumption is overestimated and conversely underestimated. Moreover, to apply the maximum flow problem a proportion of the missing outflow (due to underestimate or leak) is smoothed through all the estimated nodes to obtain inflow equal to outflow. We could detect through connected sub-graph of the network (in which flow entering and leaving the area was measured) that even if sensors were saturated we still have sink node requiring outflow describing the fact that consumption was overestimated in this area. Conversely, we detected that consumption was underestimated in some area because sensors were not saturated, but all sink node were satisfied. We can therefore improve the results provided by the algorithm by spatially refining the consumption estimates by looking at connectivity of the graph.

Future work will be to enhance the methodology with more data related to management of WDN that could constrain the number of solution of the maximum flow problem with for example intervention (e.g. valve operation, pipe break) or leak detection information.

# REFERENCES

Ahuja, Ravindra K., Magnanti, Thomas L., Orlin, James B. (1993). *Network Flows: Theory, Algorithms and Applications*, Prentice Hall.

American Water Works Association. (2005). *Manual of Water Supply Practices*. PE-Design and Installation, Denver.

Auber, D. (2004). *Tulip – A huge graph visualisation framework*, Graph drawing software, Springer, Berlin, Heidelberg, pp. 105-126.

Caragliu, A., Del Bo, C., Nijkamp, P. (2011). *Smart Cities in Europe*, in Journalof Urban Technology, pp. 65--82.

Claudio,K., Couallier, V., Le Gat, Y., Saracco, J. (2014). *Water consumption estimation of an hydraulic district from a sample of users equipped with automatic meter reading*, Journal de la Société Française de Statistique, Société Française de Statistique et Société Mathématique de France, 155 (4), pp. 160-177.

Cormen, T.H., Leiserson, C.E., Rivest, R.L., Stein, C. (2009). *Introduction to Algorithms*, third edn. MIT Press, Cambridge.

Edmonds J. and Karp Richard M.,(1972) *Theoretical improvements in algorithmic efficiency for network flow problems*, Journal of the ACM, Association for Computing Machinery (ACM), vol. 19, no 2, p. 248–264

Ford, L. and Fulkerson, D. (1962). *Flows in Networks*. Princeton Univ. Press.

Goldberg, A.V., Tardos, É., and Tarjan, R.E. (1989). *Network flow algorithms*, Tech. Report STAN-CS-89-1252, Stanford University CS Dept.

Guhl, F. and Brémond, B. (2000) Optimisation du fonctionnement des réseaux d'eau potable. Prise en compte de l'aspect stochastique de la demande [Optimization of the operation of drinking water networks. Considering the stochastic aspect of demand]. Engineering - E A T, IRSTEA 2000 edition, p. 15 - p. 23.

Heineman G.T., Pollice G., Selkow S. (2008), *Chapter 8: Network Flow Algorithms. Algorithms in a Nutshell*, Oreilly Media, pp. 226-250

Obradovic, D. And Lonsdale, P. (1998). *Public Water Supply. Models, Data and Operational Management*, E&FN SPON, Routledge, London.

Perrot A., Auber D., (2015), *FATuM –Fast Animated Transitions using Multi-Buffers*. 19th international Conference "information Visualisation 2015", Jul 2015, Barcelone, Spain.

United Nations, Department of Economic and Social Affairs. (2007). *Population Division, World Population Prospects: The 2006 Revision*, Highlights, Working Paper No. ESA/P/WP.202.

Veron, J. (2007). *Population & Sociétés - La moitié de la population mondiale vit en ville* [Population & Societies - Half of the world's population lives in cities], Institut National d'Etudes Démographiques, 435.